# COMBINATORIAL METHOD OF POLYNOMIAL EXPANSION OF SYMMETRIC BOOLEAN FUNCTIONS

Danila A. Gorodecky

The United Institute of Informatics Problems of National Academy of Sciences of Belarus, Minsk, 220012, Belarus, danila.gorodecky@gmail.com.

**Abstract**

A novel polynomial expansion method of symmetric Boolean functions is described. The method is efficient for symmetric Boolean function with small set of valued numbers and has the linear complexity for elementary symmetric Boolean functions, while the complexity of the known methods for this class of functions is quadratic. The proposed method is based on the consequence of the combinatorial Lucas theorem.

**Keywords:** *polynomial expansion, symmetric Boolean function, carrier vector, reduced Zhegalkin spectrum, complexity*

## 1. Introduction

The polynomial expansion is among the most complex tasks of the discrete mathematics. The polynomial expansion can be used to define the fifty-fifty distribution of 0 and 1 in the Steinhaus triangle, to synthesize modular summators, to find an algebraic immunity in cryptography and to solve various theoretical problems and practical applications.

Because of high computational complexity of generation of the polynomial for an arbitrary Boolean function the universal methods of the polynomial expansion are not effective. Therefore the methods of generation of expansions for various classes of Boolean functions are more effective. One of these classes is symmetric Boolean functions (SBF).

It is known many methods of the polynomial expansion of SBF. One of the most effective methods is the transeunt triangle method [1]. It has the complexity $O\left(n^2\right)$. The known methods have the redundant computations, i.e. the intermediate computations should be produced to generate the polynomial expansion.

The article represents the method of the polynomial expansion with the complexity $O(n)$ in particular cases. The method could be applied to solve the task of polynomial expansion, as well as the reverse task, i.e. representation of the function described by the polynomial.

The method is based on the consequence of the combinatorial Lucas theorem, since it is referred as the combinatorial method.

## 2. Main definitions

An arbitrary Boolean function $F = F(X)$ of the $n$ variables, where $X = (x_1, x_2, ..., x_n)$, with unchanged value after swapping any couple of variables $x_i$ and $x_j$, where $i \neq j$ and $i, j = 1,2,...,n$, is called SBF.

SBF $F$ of the $n$ variables is characterised by the set of valued numbers $A(F) = \{a_1, a_2, ..., a_r\}$. The function $F$ is equal 1 if and only if the set of variables $x_1, x_2, ..., x_n$ has exactly $a_i$ numbers of 1's, where $0 \le a_i \le n$, $0 \le i \le r$ and $0 \le r \le n+1$. These SBFs are referred as $F_n^{a_1, a_2, ..., a_r}$. If $r = 1$, then a function $F = F_n^a(X)$ is called elementary SBF (ESBF).

There is one-to-one correspondence between SBF $F_n^{a_1, a_2, ..., a_r}$ and $(n+1)-$bits binary code $\pi = (\pi_0, \pi_1, ..., \pi_n)$ – the carrier vector [2] (or the reduced truth vector [3]), where the $i-$th entry $\pi_i$ is a value of the function $F$ with the $i$ numbers of 1's, where $0 \le i \le n$. In other words, $\pi_i = 1$ if and only if the $i$ is the valued number of the SBF $F$.

The following formula is true for an arbitrary SBF $F$:

$$F(X) = \bigvee_{i=0}^{n} \pi_i F_n^i(X) = \bigoplus_{i=0}^{n} \pi_i F_n^i(X). \tag{1}$$

Positive polarity Reed-Muller polynomial (all variables are uncomplemented) is called as Zhegalkin polynomial and is referred as $P(F)$.

SBF $F$ of the $n$ variables is called the polynomial-unate SBF (PUSBF or homogeneous SBF [4]), if the Zhegalkin polynomial form $P(F)$ contains $\binom{n}{i}$ $i-$rank products with the $i$ positive literals, where $0 \le i \le n$. This function referred as $F = E_n^i$. Hence it follows

$$E_n^0 = 1,$$

$$E_n^1 = x_1 \oplus x_2 \oplus \ldots \oplus x_n,$$

$$E_n^2 = x_1x_2 \oplus \ldots \oplus x_1x_n \oplus \ldots \oplus x_{n-1}x_n,$$

$$\ldots$$

$$E_n^n = x_1x_2 \ldots x_n.$$

In general case, the polynomial form $P(F)$ of SBF $F = F(X)$ can be represented as:

$$P(F) = \gamma_0 \oplus \gamma_1(x_1 \oplus x_2 \oplus \ldots \oplus x_n) \oplus$$
$$\oplus \gamma_2(x_1x_2 \oplus \ldots \oplus x_1x_n \oplus \ldots \oplus x_{n-1}x_n) \oplus \ldots \oplus \gamma_n x_1x_2 \ldots x_n,$$

where $\gamma(F) = (\gamma_0, \gamma_1, \gamma_2, \ldots, \gamma_n)$ is the reduced Zhegalkin (Reed-Muller) spectrum of SBF. It follows

$$E(X) = \bigoplus_{i=0}^{n} \gamma_i E_n^i(X). \tag{2}$$

From the other hand PUSBF $F$ of the $n$ variables is characterised by the set of polynomial numbers $B(E) = \{b_1, b_2, \ldots, b_q\}$. The $j-$th entry of the reduced Zhegalkin spectrum $\gamma(E) = (\gamma_0, \gamma_1, \ldots, \gamma_n)$ is equal 1 if and only if $b_j = 1$, where $0 \le j \le q$ and $0 \le q \le n$. If $q = 1$, then a function $E_n^b$ is called the elementary PUSBF (EPUSBF).

The article provides the method of the transformation of the reduced truth vector $\pi(F)$ to the reduced spectrum $\gamma(F)$, i.e. $\gamma\left(E_n^{b_1, b_2, \ldots, b_q}\right)$ to $\pi\left(E_n^{b_1, b_2, \ldots, b_q}\right)$, and backwards, i.e. $\pi\left(F_n^{a_1, a_2, \ldots, a_r}\right)$ to $\gamma\left(F_n^{a_1, a_2, \ldots, a_r}\right)$.

### 3. Combinatorial method of generation of the carrier vector

The combinatorial method of the generating of the reduced truth vector $\pi\left(E_n^{b_1, b_2, \ldots, b_q}\right)$ and the reduced spectrum $\gamma\left(F_n^{a_1, a_2, \ldots, a_r}\right)$ is proposed below.

#### 3.1. Generation of the carrier vector $\pi\left(E_n^b\right)$

The process of the generating of the carrier vector $\pi\left(E_n^b\right)$ of the EPUSBF $E_n^b$ could be demonstrated on the example.

**Example 1.** Let's assume that it is necessary to get the carrier vector $\pi=(\pi_0,\pi_1,...,\pi_6)$ for the PUSBF $F(X)=E_6^2(X)$.

From the condition it follows that $\gamma=(0,0,1,0,0,0,0)$ and

$$
\begin{array}{lccccccccccc}
 & 1 & & 2 & & 3 & & 4 & & 5 & \\
E_6^2 = & x_1x_2 & \oplus & x_1x_3 & \oplus & x_1x_4 & \oplus & x_1x_5 & \oplus & x_1x_6 & \oplus \\
 & & \oplus & x_2x_3 & \oplus & x_2x_4 & \oplus & x_2x_5 & \oplus & x_2x_6 & \oplus \\
 & & & & \oplus & x_3x_4 & \oplus & x_3x_5 & \oplus & x_3x_6 & \oplus \\
 & & & & & & \oplus & x_4x_5 & \oplus & x_4x_6 & \oplus \\
 & & & & & & & & \oplus & x_5x_6 . &
\end{array}
\qquad (3)
$$

Note, that number of the column is equal to the number of factors in the column which are included in the polynomial of the function $E_6^2(X)$. The polynomial $P\left(E_6^2\right)$ contains $\binom{6}{2}=15$ $2-$rank products.

To generate the carrier vector $\pi=(\pi_0,\pi_1,\pi_2,\pi_3,\pi_4,\pi_5,\pi_6)$ the $i-$th entry $\pi_i$ should be defined with the following arguments, where $i=\overline{0,6}$:

– assuming $\pi_0=1$ then $E_6^2$ contains $F_6^0$. But it is impossible, because the polynomial (3) of $E_6^2(X)$ doesn't contain term 1. In this case $E_6^2=0$ and therefore $\pi_0=0$;

– assuming $\pi_1=1$ then $E_6^2$ contains $F_6^1$. According to the definition of the ESBF $F_6^1=1 \Leftrightarrow$ polynomial (3) is equal 1 for $x_1=1$ and $x_2=x_3=...=x_6=0$. But it is impossible. In this case $E_6^2=0$ and therefore $\pi_1=0$;

– assuming $\pi_2=1$ then $E_6^2$ contains $F_6^2$. According to the definition of the ESBF $F_6^2=1 \Leftrightarrow$ polynomial (3) is equal 1 for $x_1=x_2=1$ and $x_3=x_4=x_5=x_6=0$. Thus the only factor from the first column of the polynomial (3) is equal 1. In this case $E_6^2=1$ and therefore $\pi_2=1$;

– assuming $\pi_3=1$ then $E_6^2$ contains $F_6^2\oplus F_6^3$. According to the definition of the ESBF $F_6^3=1$ $\Leftrightarrow$ polynomial (3) is equal 1 for $x_1=x_2=x_3=1$ and $x_4=x_5=x_6=0$. Thus the factors from the first and second columns of the polynomial (3) are equal 1. Since the number of the unity components is the odd number, then in this case $E_6^2=1$ and therefore $\pi_3=1$;

– assuming $\pi_4=1$ then $E_6^2$ contains $F_6^2\oplus F_6^3\oplus F_6^4$. According to the definition of the ESBF $F_6^4=1 \Leftrightarrow$ polynomial (3) is equal 1 for $x_1=x_2=x_3=x_4=1$ and $x_5=x_6=0$. Since the factors from the first, second and third columns of the polynomial (3) are equal 1, then the number of the unity components is the even number. In this case $E_6^2=0$ and therefore $\pi_4=0$;

– assuming $\pi_5=1$ then $E_6^2$ contains $F_6^2\oplus F_6^3\oplus F_6^5$. According to the definition of the ESBF $F_6^5=1 \Leftrightarrow$ polynomial (3) is equal 1 for $x_1=x_2=x_3=x_4=x_5=1$ and $x_6=0$. Since the factors from the first, second, third and fourth columns of the polynomial (3) are equal 1, then the number of the unity components is the even number. In this case $E_6^2=0$ and therefore $\pi_5=0$;

– assuming $\pi_6 = 1$ then $E_6^2$ contains $F_6^2 \oplus F_6^3 \oplus F_6^6$. According to the definition of the ESBF $F_6^6 = 1 \Leftrightarrow$ polynomial (3) is equal 1 for $x_1 = x_2 = x_3 = x_4 = x_5 = x_6 = 1$. Since the factors from all columns of the polynomial (3) are equal 1, thrn the number of the unity components is the odd number. In this case $E_6^2 = 1$ and therefore $\pi_6 = 1$.

As the result the carrier vector of the EPUSBF $E_6^2$ is $\pi(E_6^2) = (0,0,1,1,0,0,1)$ and $P(E_6^2) = F_6^2 \oplus F_6^3 \oplus F_6^6$.

It is worth to pay attention to the fact that the value of the polynomial depends only on the parity number of unity factors.

The reasoning used in the example 1 may be summarized with the theorem.

**Theorem 1.** The $i-$th entry $\pi_i$ of the carrier vector $\pi = (\pi_0, \pi_1, ..., \pi_n)$ of the PUSBF $E_n^b = E_n^b(x_1, x_2, \dots, x_n)$ is calculated by using the formula:

$$\pi_i = \begin{cases} 1, \ if \ \begin{pmatrix} i \\ b \end{pmatrix} = 1 \ (\mathrm{mod} \ 2); \\ 0 - otherwise, \end{cases} \tag{4}$$

where $i = \overline{b, n}$.

*Proof.* Let’s consider three cases of relations $i$ and $b$.

The first case where $i < b$. Then the number of the unity terms $i$ is less then the $b-$rank products and $E_i^b = 0$ (see the first and second cases of the example 1). Therefore $\pi_i = 0$.

The second case where $i = b$, i.e. $x_1 = x_2 = ... = x_b = 1$ and $x_{b+1} = x_{b+2} = ... = x_n = 0$. Thus just one $b-$rank term of the PUSBF $E_b^b$ is equal 1 and $P(E_b^b) = x_1 x_2 ... x_b = 1$ (see the third case of the example 1). In this case $\pi_i = 1$.

The third case where $i > b$, i.e. $x_1 = x_2 = ... = x_b = ... = x_i = 1$ and $x_{i+1} = ... = x_n = 0$. Thus the function $E_i^b$ is represented by the polynomial $P(E_i^b) = \underbrace{x_1 x_2 ... x_b \oplus x_1 x_2 ... x_{b-1} x_{b+1} \oplus ... \oplus x_1 x_2 ... x_{b-1} x_i \oplus ... \oplus x_{i-b+1} x_{i-b+2} ... x_i}_{i}$. Since the value of $P(E_i^b)$ is determined by even-odd of $i$. Thus $E_i^b = \begin{cases} 1, \ if \ \begin{pmatrix} i \\ b \end{pmatrix} = 1 \ (\mathrm{mod} \ 2); \\ 0 - otherwise. \end{cases}$ In this case

$$\pi_i = \begin{cases} 1, \ if \ \begin{pmatrix} i \\ b \end{pmatrix} = 1 \ (\mathrm{mod} \ 2); \\ 0 - otherwise. \end{cases}$$

The statement of the theorem is proved.

As a result of *Theorem 1* the carrier vector $\pi$ of the PUSBF $E_n^b$ corresponds to the following form

$$\pi = \left( \underbrace{0, 0, ..., 0}_{b}, 1, \underbrace{\pi_{b+1}, ..., \pi_n}_{n-b} \right). \tag{5}$$

Consequence of the Lucas theorem is helpful for calculation using the formula (4). It determines the even-odd of the number $\binom{i}{b}$ and as follows.

**Theorem 2. (*Consequence of the Lucas theorem*) [5].** The number $\binom{n}{b} = 1 \pmod 2 \Leftrightarrow$ each bit of $b$ is no more than the same bit of $n$, where $n \geq b$ in decimal representation.

Note, that the binary length $n = (n_{\delta_1}, n_{\delta_1 - 1}, ..., n_1)$ and $b = (b_{\delta_2}, b_{\delta_2 - 1}, ..., b_1)$ is defined as $\delta_1 = [\log_2 n] + 1$ and $\delta_2 = [\log_2 b] + 1$ respectively.

**Example 2.** Let's define even-odd of the number $\binom{n}{b}$ using *Theorem 2* and assuming $n = 11$ and $b$ for two cases a) $b = 2$ and b) $b = 5$.

The length of $n$ is $\delta_1 = [\log_2 11] + 1 = 4$, then $n = (n_4, n_3, n_2, n_1) = (1010)$, and

a) $\delta_2 = [\log_2 2] + 1 = 2$, then $b = (b_2, b_1) = (10)$;

b) $\delta_2 = [\log_2 5] + 1 = 3$, then $b = (b_3, b_2, b_1) = (101)$.

For the case a) the binary representations $b$ and $n$ are comparable and satisfy the condition of Theorem 2, as pictured in Figure 1 a).

For the case b) the binary representations of $b$ and $n$ are not comparable and do not satisfy the condition of Theorem 2, as pictured in Figure 1 b).

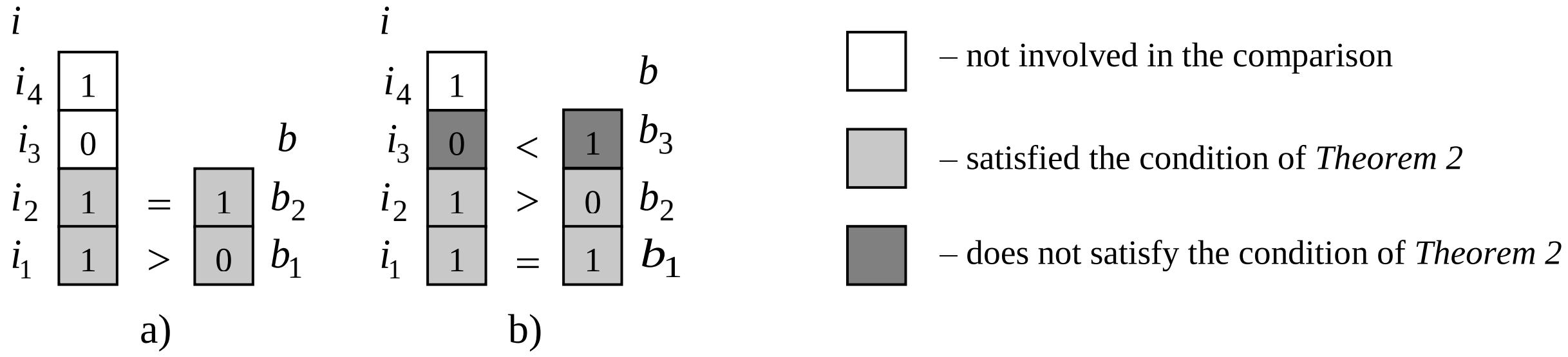

Figure 1. Definition of even-odd of the numbers a) $\binom{11}{2}$; b) $\binom{11}{5}$

As a result, in the case a) the number $\binom{11}{2}$ is the odd and thus $\binom{11}{2} = 1 \pmod 2$; in the case b) the number $\binom{11}{5}$ is the even and thus $\binom{11}{5} = 0 \pmod 2$.

Let's generate the carrier vector for the function in above example 1 using *Theorem 1* and *Theorem 2*.

**Example 3.** Let's assume that it is necessary to generate $\pi(E_6^2)$.

From the condition it follows $\gamma(E_6^2) = (0, 0, 1, 0, 0, 0, 0)$. According to the formula (5) $\pi_0 = \pi_1 = 0$, $\pi_2 = 1$ and $\pi = (0, 0, 1, \pi_3, \pi_4, \pi_5, \pi_6)$.

Therefore, in order to find $\pi_3, \pi_4, \pi_5, \pi_6$ the even-odd order of the Binomial coefficients $\binom{3}{2}$, $\binom{4}{2}$, $\binom{5}{2}$, $\binom{6}{2}$ respectively should be defined. From the *Theorem 2* they could be defined as shown on figure 2.

The figure 2 is analogous to representation as follows: $\binom{3}{2}=\binom{11}{10} \Rightarrow \pi_3=1$; $\binom{4}{2}=\binom{100}{10} \Rightarrow \pi_4=0$; $\binom{5}{2}=\binom{101}{10} \Rightarrow \pi_5=0$; $\binom{6}{2}=\binom{110}{10} \Rightarrow \pi_6=1$.

As the result the carrier vector is $\pi\left(E_6^2\right)=(0,0,1,1,0,0,1)$ and $P\left(E_6^2\right)=F_6^2 \oplus F_6^3 \oplus F_6^6$.

The procedure of calculating of the entries of the carrier vector using consequence of the Lucas theorem is called combinatorial methods.

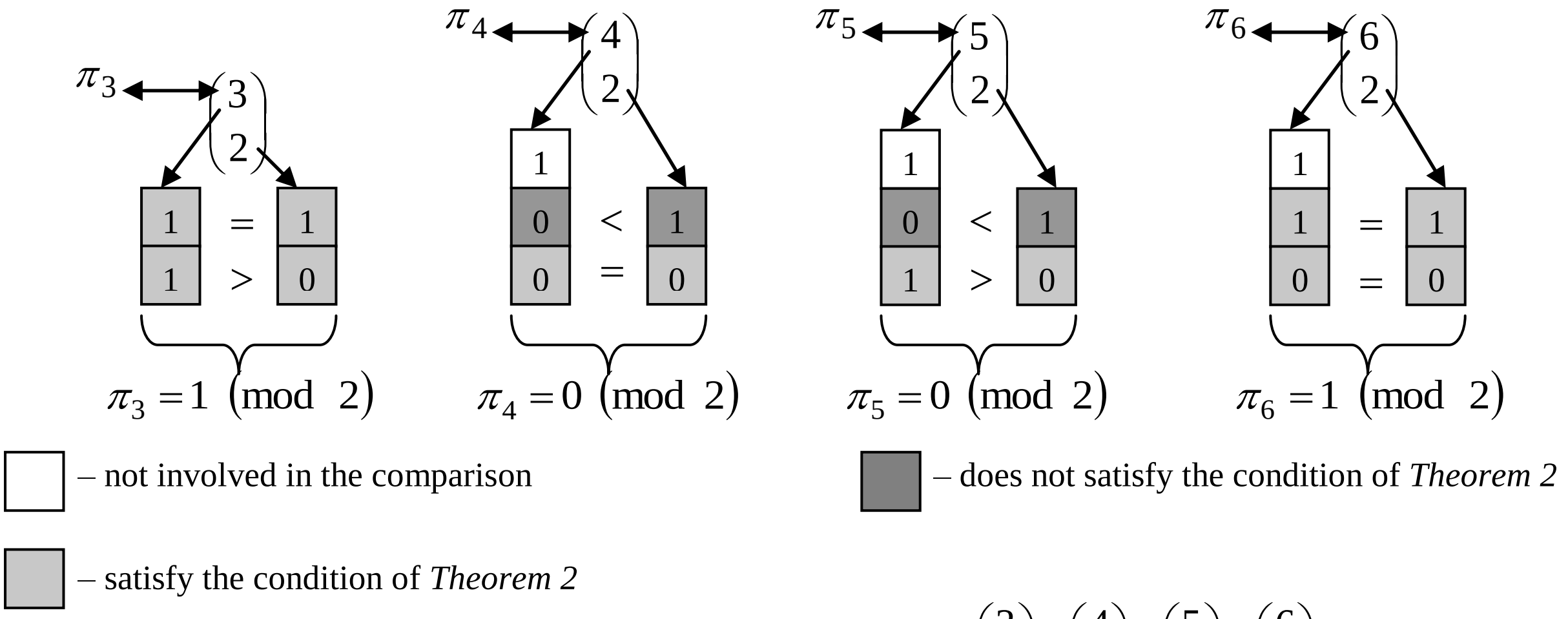

Figure 2. Definition of even-odd of the numbers $\binom{3}{2}$, $\binom{4}{2}$, $\binom{5}{2}$, $\binom{6}{2}$

### 3.2. Generation of the carrier vector $\pi\left(E_n^{b_1,b_2,\ldots,b_q}\right)$

The combinatorial method of generating of the carrier vector for the EPUSBF $E_n^b=E_n^b(x_1,x_2,\ldots,x_n)$ can be generalized for an arbitrary PUSBF $E_n^{b_1,b_2,\ldots,b_q}=E_n^{b_1,b_2,\ldots,b_q}(x_1,x_2,\ldots,x_n)$ with the following theorem.

**Theorem 3.** The $i-$th entry $\pi_i$ of the carrier vector $\pi\left(E_n^{b_1,b_2,\ldots,b_q}\right)=(\pi_0,\pi_1,\ldots,\pi_n)$ of the PUSBF $E_n^{b_1,b_2,\ldots,b_q}$ is calculated with the following formula:

$$\pi_i=\begin{cases}1, \ if \ \binom{i}{b_1}+\binom{i}{b_2}+\ldots+\binom{i}{b_q}=1 \pmod 2; \\ 0 \ - \ otherwise,\end{cases} \tag{6}$$

where $i=\overline{b_1+1,n}$. Note, that $\binom{i}{b_j}$ for $i<b_j$ is meaningless, where $j=\overline{1,q}$, therefore let's assume $\binom{i}{b_j}=0$ for $i<b_j$.

The proof of *Theorem 3* follows from *Theorem 1*.

**Example 4.** Let's assume that it is necessary to generate $\pi\left(E_{10}^{5,7,8}\right)$.

From the condition it follows $\gamma\left(E_{10}^{5,7,8}\right)=(0,0,0,0,0,1,0,1,1,0,0)$. According to the formula (5) it follows $\pi_0=\pi_1=\pi_2=\pi_3=\pi_4=0$ and $\pi_5=1$. Thus $\pi\left(E_{10}^{5,7,8}\right)=(0,0,0,0,0,1,\pi_6,\pi_7,\pi_8,\pi_9,\pi_{10})$.

According to the formula (6) and *Theorem 2* it is easy to define $\pi_6, \pi_7, \pi_8, \pi_9$ and $\pi_{10}$ as shown on the figure 3.

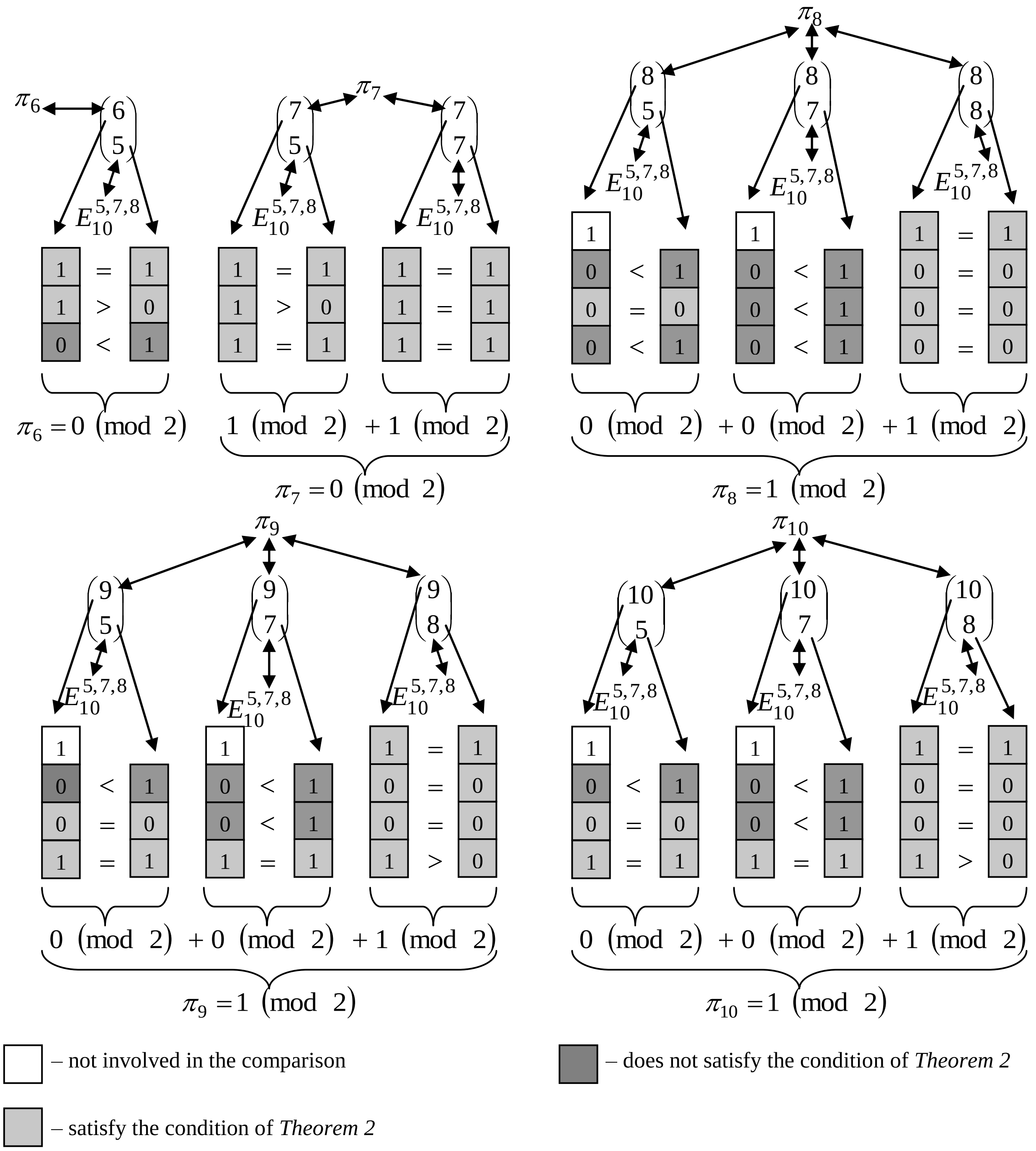

Figure 3. Calculating of the $\pi_6, \pi_7, \pi_8, \pi_9, \pi_{10}$ for $\pi\left(E_{10}^{5,7,8}\right)$

The figure 3 is analogous to representation as follows:

$$\binom{6}{5} = \binom{110}{101} = 0 \pmod 2 \Rightarrow \pi_6 = 0;$$

$$\binom{7}{5} + \binom{7}{7} = \binom{111}{101} + \binom{111}{111} = 1 \pmod 2 + 1 \pmod 2 = 0 \pmod 2 \Rightarrow \pi_7 = 0;$$

$$\binom{8}{5}+\binom{8}{7}+\binom{8}{8}=\binom{1000}{101}+\binom{1000}{111}+\binom{1000}{1000}=0(\mathrm{mod}\ 2)+0(\mathrm{mod}\ 2)+1(\mathrm{mod}\ 2)=1(\mathrm{mod}\ 2)\Rightarrow$$

$\Rightarrow \pi_8=1;$

$$\binom{9}{5}+\binom{9}{7}+\binom{9}{8}=\binom{1001}{101}+\binom{1001}{111}+\binom{1001}{1000}=0\ (\mathrm{mod}\ 2)+0\ (\mathrm{mod}\ 2)+1\ (\mathrm{mod}\ 2)=1\ (\mathrm{mod}\ 2)$$

$\Rightarrow \pi_9=1;$

$$\binom{10}{5}+\binom{10}{7}+\binom{10}{8}=\binom{1010}{101}+\binom{1010}{111}+\binom{1010}{1000}=0(\mathrm{mod}\ 2)+0(\mathrm{mod}\ 2)+1(\mathrm{mod}\ 2)=1(\mathrm{mod}\ 2)$$

$\Rightarrow \pi_{10}=1.$

Thus $\pi_6=\pi_7=0$ and $\pi_8=\pi_9=\pi_{10}=1$. As the result the carrier vector of the PUSBF $E_{10}^{5,7,8}$ is $\pi\left(E_{10}^{5,7,8}\right)=(0,0,0,0,0,1,0,0,1,1,1)$ and $P\left(E_{10}^{5,7,8}\right)=F_{10}^{5}\oplus F_{10}^{7}\oplus F_{10}^{8}$.

## 4. Generation of the reduced spectrum $\gamma\left(E_n^{b_1,b_2,\ldots,b_q}\right)$

The combinatorial method of the generation of the carrier vector $\pi\left(F_n^{a_1,a_2,\ldots,a_r}\right)$ can be applied to the generation of the reduced spectrum $\gamma\left(F_n^{a_1,a_2,\ldots,a_r}\right)$, where $F_n^{a_1,a_2,\ldots,a_r}$ is the SBF.

To solve the task of the generating of the reduced spectrum $\gamma\left(F_n^{a_1,a_2,\ldots,a_r}\right)$ *Theorem 1* and *Theorem 3* can be adapted to the two following forms.

**Theorem 4.** The $i-$th entry $\gamma_i$ of the reduced spectrum $\gamma=(\gamma_0,\gamma_1,\ldots,\gamma_n)$ of the PUSBF $F_n^a=F_n^a(x_1,x_2,\ \ldots\ ,x_n)$ is calculated with the following formula:

$$\gamma_i=\begin{cases}1,\ \mathit{if}\ \binom{i}{a}=1\ (\mathrm{mod}\ 2);\\ 0\ -\ \mathit{otherwise},\end{cases} \qquad (7)$$

where $i=\overline{a,n}$.

**Theorem 5.** The $i-$th entry $\gamma_i$ of the reduced spectrum $\gamma\left(F_n^{a_1,a_2,\ldots,a_r}\right)=(\gamma_0,\gamma_1,\ldots,\gamma_n)$ of the SBF $F_n^{a_1,a_2,\ldots,a_r}$ is calculated with the following formula:

$$\gamma_i=\begin{cases}1,\ \mathit{if}\ \binom{i}{a_1}+\binom{i}{a_2}+\ldots+\binom{i}{a_r}=1\ (\mathrm{mod}\ 2);\\ 0\ -\ \mathit{otherwise},\end{cases} \qquad (8)$$

where $i=\overline{a_1+1,n}$. Note, that $\binom{i}{a_j}$ for $i<a_j$ is meaningless, where $j=\overline{1,r}$, therefore let's assume $\binom{i}{a_r}=0$ and $i<a_j$.

According to *Theorem 4* and *Theorem 5* the reduced spectrum $\gamma=(\gamma_0,\gamma_1,\ldots,\gamma_n)$ of the SBF $F_n^{a_1,a_2,\ldots,a_r}$ corresponds to the following form

$$\gamma = \left( \underbrace{0,0,...,0}_{a},1,\underbrace{\gamma_{a+1},...,\gamma_n}_{n-a} \right). \quad (9)$$

The example of the application of *Theorem 5* will be considered.

**Example 5.** Let's generate the reduced spectrum $\gamma\left(F_7^{2,3}\right)$.

From the condition it follows the carrier vector is $\pi\left(E_7^{2,3}\right) = (0,0,1,1,0,0,0,0)$. According to the formula (9) it follows $\gamma_0 = \gamma_1 = 0$, $\gamma_2 = 1$ and $\gamma\left(F_7^{2,3}\right) = (0,0,1,\gamma_3,\gamma_4,\gamma_5,\gamma_6,\gamma_7)$.

According to the formula (8) and *Theorem 5* it is easy to define $\gamma_3,\gamma_4,\gamma_5,\gamma_6$ and $\gamma_7$ as shown below:

$$\binom{3}{2}+\binom{3}{3}=\binom{11}{10}+\binom{11}{11}=1 \pmod 2 + 1 \pmod 2 = 0 \pmod 2 \Rightarrow \gamma_3 = 0;$$

$$\binom{4}{2}+\binom{4}{3}=\binom{100}{10}+\binom{100}{11}=0 \pmod 2 + 0 \pmod 2 = 0 \pmod 2 \Rightarrow \gamma_4 = 0;$$

$$\binom{5}{2}+\binom{5}{3}=\binom{101}{10}+\binom{101}{11}=0 \pmod 2 + 0 \pmod 2 = 0 \pmod 2 \Rightarrow \gamma_5 = 1;$$

$$\binom{6}{2}+\binom{6}{3}=\binom{110}{10}+\binom{110}{11}=1 \pmod 2 + 0 \pmod 2 = 1 \pmod 2 \Rightarrow \gamma_6 = 1;$$

$$\binom{7}{2}+\binom{7}{3}=\binom{111}{10}+\binom{111}{11}=1 \pmod 2 + 1 \pmod 2 = 0 \pmod 2 \Rightarrow \gamma_7 = 0.$$

As the result the reduced spectrum of the function $F_7^{2,3}$ is $\gamma\left(F_7^{2,3}\right) = (0,0,1,0,0,1,1,0)$ and $P\left(F_7^{2,3}\right) = E_7^2 \oplus E_7^5 \oplus E_7^6$.

## 5. The complexity of the combinatorial method

The complexity of the proposed method can be defined as the number of the binary operations XOR (or OR) and is referred $S_1$ for the EPUSBF $E_n^b$ (or ESBF $F_n^a$) and $S_2$ for the PUSBF $E_n^{b_1,b_2,...,b_q}$ (or SBF $F_n^{a_1,a_2,...,a_r}$ ).

The positive relationship of two binary vectors is $(x_t, x_{t-1},..., x_1) \geq (y_t, y_{t-1},..., y_1)$, if $x_i \geq y_i$, where $i = \overline{1,t}$. In this way to define the relationship $x_i \geq y_i$ the following condition should be satisfied

$$x_i \vee \overline{y_i} = 1. \quad (10)$$

Therefore according to the condition of *Theorem 2* the complexity (the number of operations (10)) of the computation of the number $\binom{i}{b} = 1 \pmod 2$ is $[\log_2 b] + 1$. From *Theorem 1* it follows the complexity of the computation of the carrier vector $\pi\left(E_n^b\right) = (\pi_0, \pi_1,..., \pi_n)$ is

$$S_1 = ([\log_2 b] + 1) \cdot (n - b). \quad (11)$$

From *Theorem 3* it follows the complexity of the computation of the carrier vector $\pi\left(E_n^{b_1,b_2,...,b_q}\right) = (\pi_0, \pi_1,..., \pi_n)$ is

$$S_2 = ([\log_2 b_1]+1)\cdot(n-b_1)+([\log_2 b_2]+1)\cdot(n-b_2+1)+([\log_2 b_2]+1)\cdot(n-b_2+1)+\ldots+$$
$$+([\log_2 b_q]+1)\cdot(n-bq)+(n-b_2+1)+(n-b_3+1)+\ldots+(n-b_q+1)=$$
$$=([\log_2 b_1]+1)\cdot(n-b_1)+\sum_{i=2}^{q}([\log_2 b_i]+1)\cdot(n-b_i+1)+\sum_{i=2}^{q}(n-b_i+1)= \qquad (12)$$
$$=([\log_2 b_1]+1)\cdot(n-b_1)+\sum_{i=2}^{q}([\log_2 b_i]+1)\cdot(n-b_i+1).$$

The complexity of the calculation of $\gamma(F_n^a)$ and $\gamma(F_n^{a1,a2,\ldots,ar})$ can be calculated with (11) and (12) respectively.

**6. Discussion**

There are some effective methods to solve the task of polynomial expansion, i.e. generating of the reduced spectrum $\gamma(F_n^{a_1,a_2,\ldots,a_r})=(\gamma_0,\gamma_1,\ldots,\gamma_n)$ and the reverse task of generating of the carrier vector $\pi(E_n^{b_1,b_2,\ldots,b_r})=(\pi_0,\pi_1,\ldots,\pi_n)$. One of these methods is the transeunt triangle method. It was originally proposed by V.P. Suprun for SBF [1]. Then the method was generalized for arbitrary Boolean functions [6]. The transeunt triangle method is the most effective method for polynomial expansion of the SBF $F=F(x_1,x_2,\ldots,x_n)$ and has the complexity $O(n^2)$.

The transeunt triangle form is generated from the upper base to the bottom using the XOR-operation (see example 6). Thus the number of XOR-operations defines the complexity $S_T$ of the transeunt triangle method and is

$$S_T = \frac{n^2+n}{2}. \qquad (13)$$

There is the example of the implementation of the transeunt triangle method for generating of the carrier vector $\pi(E_n^b)=(\pi_0,\pi_1,\ldots,\pi_n)$.

**Example 6.** Let's assume the reduced spectrum $\gamma(E_6^2)=(0,0,1,0,0,0,0)$, i.e. $n=6$ and $b=2$. In order to generate $\pi(E_6^2)$ the transeunt triangle method will be used.

From the condition the upper base of the triangle will $\gamma(E_6^2)=(0,0,1,0,0,0,0)$ and it takes the form as follow:

```
0   0   1   0   0   0   0
  0   1   1   0   0   0
    1   0   1   0   0
      1   1   1   0
        0   0   1
          0   1
            1
```

According to the transeunt triangle method the left side of the triangle corresponds to the reduced carrier vector and it is equal to $\pi(E_6^2)=(0,0,1,1,0,0,1)$. Therefore $E_6^2=F_6^2\oplus F_6^3\oplus F_6^6$.

Using the formula (13) the complexity of the computation of the $\pi\left(E_6^2\right)$ with the transeunt triangle method is $S_T = 15$. On the other side, to complexity of performing the same task using the combinatorial method, according to the formula (11) and as shown in example 3, is $S_1 = 8$.

Firstly let's compare the complexity $S_1$ (formula (11)) of the combinatorial method proposed in the article and the complexity $S_T$ (formula (13)) of the transeunt triangle method for SBF $E_n^a$ (or $F_n^a$). The illustration of the comparison of $S_1$ and $S_T$ is shown on the Figure 4.

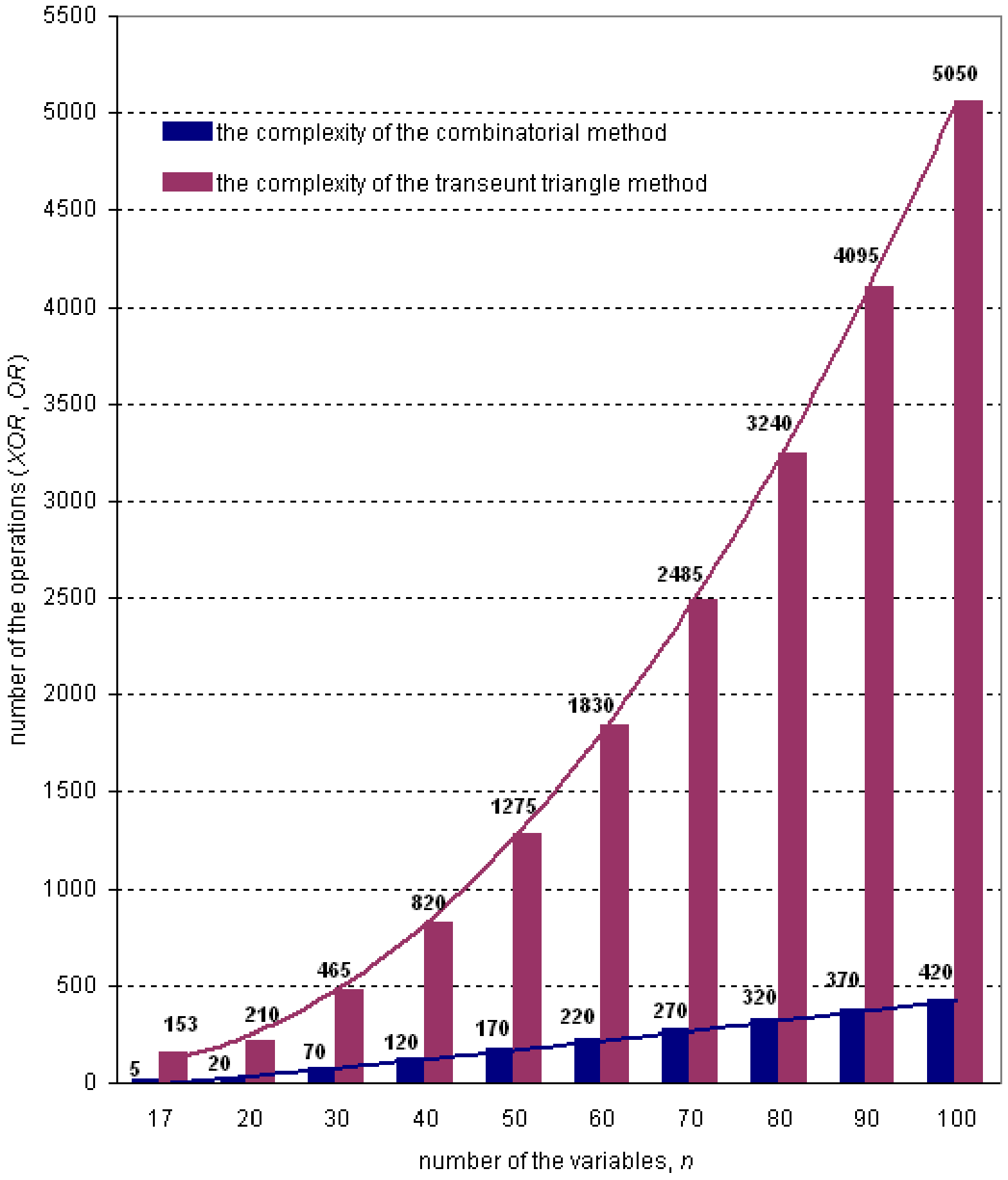

**Figure 4.** The comparison of the complexity $S_1$ of the combinatorial method and the complexity $S_T$ of the transeunt triangle method

As it can be seen at figure 4 the combinatorial method for EPUSBF has the linear complexity. The proposed method can provide ten times more efficiency for 80 variables in comparison with the transeunt triangle method. The complexity of the combinatorial method (formula (11)) is calculated for the worst case, i.e. for the EPUSBF $E_n^b$, where $b = 16$.

The complexity of the combinatorial method in comparison with the complexity of the transeunt triangle method for PUSBF $E_n^{b_1, b_2, \ldots, b_r}$, where $r > 1$, strongly depends on numbers included in the set of the polynomial numbers. As a result, the table demonstrates the threshold of the efficiency of the combinatorial method in comparison with the transeunt triangle method.

The second column contains the power set of the polynomial numbers $B(E)$ for which the complexity $S_2$ and $S_T$ is approximately equal. The third column contains the set of the polynomial numbers $B(E)$ for which the complexities of both methods are the same. Any other set of the polynomial numbers $B(E)$ provides a lower complexity of the combinatorial method for the number of the variables specified in the first column. The fourth column shows the ratio of the set of the polynomial numbers to all variables specified in the first column. Two right columns show the comparable complexities of the combinatorial and the transeunt triangle methods.

Table
The efficiency of the combinatorial method and the transeunt triangle method

| Number of the variables $n$ that the function $E_n^{b_1,b_2,\ldots,b_r}$ depends on | Number $r$ of the polynomial numbers $b_1,b_2,\ldots,b_r$ for $S_2 \approx S_T$ | Set of the polynomial numbers $B(E)=b_1,b_2,\ldots,b_r$ | Percentage of the number $r$ of the variables $n$, % | Complexity of the combinatorial method, $S_2$ | Complexity of the transeunt triangle method, $S_T$ |
|---|---|---|---|---|---|
| 10 | 3 | $\{2,3,4\}$ | 30 | 53 | 55 |
| 20 | 5 | $\{4,\ldots,8\}$ | 25 | 235 | 210 |
| 30 | 6 | $\{4,\ldots,9\}$ | 20 | 483 | 465 |
| 40 | 7 | $\{8,\ldots,14\}$ | 18 | 836 | 820 |
| 50 | 8 | $\{11,\ldots,18\}$ | 16 | 1266 | 1275 |
| 60 | 9 | $\{16,\ldots,24\}$ | 15 | 1840 | 1830 |
| 70 | 10 | $\{16,\ldots,25\}$ | 14 | 2520 | 2485 |
| 80 | 11 | $\{16,\ldots,26\}$ | 14 | 3295 | 3240 |
| 90 | 12 | $\{16,\ldots,27\}$ | 13 | 4765 | 4095 |
| 100 | 13 | $\{16,\ldots,28\}$ | 13 | 5130 | 5050 |
| 255 | 26 | $\{32,\ldots,57\}$ | 10 | 32988 | 32640 |
| 511 | 44 | $\{64,\ldots,107\}$ | 9 | 131355 | 130816 |
| 1023 | 76 | $\{128,\ldots,203\}$ | 7 | 521960 | 523776 |
| 2047 | 135 | $\{256,\ldots,390\}$ | 7 | 2095866 | 2096128 |
| 4095 | 242 | $\{512,\ldots,753\}$ | 6 | 8381660 | 8386560 |

## 7. Conclusions

The combinatorial method is a new method of generating of the carrier vector $\pi\left(E_n^{b_1,b_2,\ldots,b_q}\right)$ and the reduced spectrum $\gamma\left(F_n^{a_1,a_2,\ldots,a_q}\right)$ of SBF, i.e. polynomial expansion of SBF.

The proposed method is the linear complexity and the complexity of the known methods (for example, transeunt triangle method) is quadratic for EPUSBF (or ESBF). The combinatorial method provides high efficiency for small number of variables for PUSBF (or SBF).